\documentclass[pre,aps,preprint,footinbib,superscriptaddress,longbibliography,floatfix]{revtex4-1}
\usepackage{graphicx}
\usepackage{dcolumn}
\usepackage{bm}
\usepackage{amssymb}
\usepackage{pifont}
\usepackage{amsmath}
\usepackage{times}
\usepackage[nooneline]{subfigure}

\usepackage{graphicx,psfrag}

\begin{document}

\title{Firing rate equations require a spike synchrony mechanism to 
correctly describe fast oscillations in inhibitory networks}
\author{Federico Devalle}
\affiliation{Center for Brain and Cognition. Department of Information and Communication Technologies,
Universitat Pompeu Fabra, 08018 Barcelona, Spain}
\affiliation{Department of Physics, Lancaster University, Lancaster LA1 4YB, United Kingdom}
\author{Alex Roxin}
\affiliation{Centre de Recerca Matem\`atica, Campus de Bellaterra, Edifici C,  08193 Bellaterra, Barcelona, Spain.}
\author{Ernest Montbri\'o}
\affiliation{Center for Brain and Cognition. Department of Information and Communication Technologies,
Universitat Pompeu Fabra, 08018 Barcelona, Spain}

\begin{abstract}
Recurrently coupled networks of inhibitory neurons robustly generate
oscillations in the gamma band.  Nonetheless, the corresponding
Wilson-Cowan  type firing rate equation for such an inhibitory
population does not generate such oscillations without an explicit time
delay. We show that this discrepancy is due to a voltage-dependent
spike-synchronization mechanism inherent in networks of spiking
neurons which is not captured by standard firing rate equations.  Here
we investigate an exact low-dimensional description for a network of
heterogeneous canonical Class 1 inhibitory neurons which includes the
sub-threshold dynamics crucial for generating synchronous states.  In
the limit of slow synaptic kinetics the spike-synchrony mechanism is
suppressed and the standard Wilson-Cowan equations are formally
recovered as long as external inputs are also slow.  However, even in
this limit synchronous spiking can be elicited by inputs which
fluctuate on a time-scale of the membrane time-constant of the
neurons.   Our meanfield equations therefore represent an extension of the 
standard Wilson-Cowan equations in which spike synchrony is also correctly described.
\end{abstract}

\maketitle

\section{Introduction}

Since the seminal work of Wilson and Cowan \cite{WC72}, population
models of neuronal activity have become a standard tool of analysis in
computational  neuroscience. Rather than focus on the microscopic
dynamics of neurons,  these models describe the collective properties
of large numbers of neurons, typically in terms of the mean firing
rate of a neuronal ensemble. 
In general, such population models, often called firing rate equations, 
cannot be exactly  derived from the equations
of a network of spiking neurons,  but are obtained using heuristic
mean-field arguments,  see e.g.~\cite{ET10,GKN+14,DA01,Cow14,CGP14}.
Despite their heuristic nature, heuristic firing rate equations (which we call H-FRE) 
often show remarkable qualitative
agreement with the dynamics in equivalent networks of spiking neurons
~\cite{LRN+00,SHS03,RBH05,RM11}, and constitute an extremely useful
modeling tool, see e.g.
~\cite{WC73,Ama74,Nun74,EC79,BLS95,PBS+96,HS98,TPM98,Wil99,TSO+00,BCG+01,LTG+02,HT06,MRR07,MBT08,TWC+11,MR13,TDD14}.
Nonetheless, this agreement can break down once a significant fraction
of the  neurons in the population fires spikes synchronously, see
e.g.~\cite{SOA13}.  Such synchronous firing may come about due to
external drive,  but also occurs to some degree during spontaneously
generated network states. 
 
As a case in point, here we focus on partially synchronized states in
networks of heterogeneous inhibitory
neurons. Inhibitory networks are able to
generate robust  macroscopic oscillations due to the interplay of
external excitatory inputs with the inhibitory mean field produced by
the population itself.  Fast synaptic processing coupled with
subthreshold integration of inputs  introduces an effective delay in
the negative feedback facilitating the emergence of what is often called
Inter-Neuronal Gamma (ING)
oscillations~\cite{WB96,WTJ95,WCR+98,WTK+00,TJ00,BH06,BH08,BVJ07,Wan10}.  
Modeling studies with networks of spiking neurons demonstrate that, in
heterogeneous inhibitory networks,  large fractions of neurons become
frequency-entrained during these oscillatory episodes, and that the
oscillations persist for weak levels of
heterogeneity~\cite{WB96,WCR+98,TJ00}.  Traditional H-FRE 
(also referred to as Wilson-Cowan equations)
fail to describe such fast oscillations.
To overcome this limitation, explicit fixed
time delays have been considered in H-FRE 
 as a heuristic  proxy for the
combined effects of synaptic and subthreshold integration~\cite{RBH05,BH08,RM11,KFR17}.  

Here we show that fast oscillations in inhibitory networks are
correctly described  by a recently derived set of exact macroscopic
equations for quadratic integrate-and-fire neurons (that we call QIF-FRE) which explicitly take into account subthreshold  integration~\cite{MPR15}.
Specifically, the QIF-FRE reveal how oscillations arise via a
voltage-dependent spike synchronization mechanism, missing in H-FRE,
as long as the recurrent synaptic kinetics are sufficiently fast. 
In the limit of
slow recurrent synaptic kinetics intrinsically generated oscillations
are suppressed, and the QIF-FRE reduce to an equation formally identical 
to the Wilson-Cowan equation for an inhibitory population. However, even in this limit, 
fast fluctuations in external inputs can drive transient spike synchrony in the network, 
and the slow synaptic approximation of the QIF-FRE breaks down.  
This suggests that, in general, a correct macroscopic description of spiking 
networks requires keeping track of 
the mean subthreshold voltage along with the mean firing rate. 

Additionally, the QIF-FRE describe the disappearance of oscillations for 
sufficiently strong heterogeneity which is 
robustly observed in simulations of spiking networks.  
Finally, we also show that the phase diagrams of 
oscillatory states found in the QIF-FRE qualitatively match 
those observed in simulations 
of populations of more biophysically inspired Wang-Buzs\'aki neurons ~\cite{WB96}.  
This shows that not only are the QIF-FRE an exact mean-field description of 
networks of heterogeneous QIF neurons, but 
that they also provide a qualitatively accurate description of dynamical states in networks of 
spiking neurons more generally, including states with significant spike synchrony.

\section{Results}

\begin{figure}[]
\centerline{\includegraphics[width=125mm,clip=true]{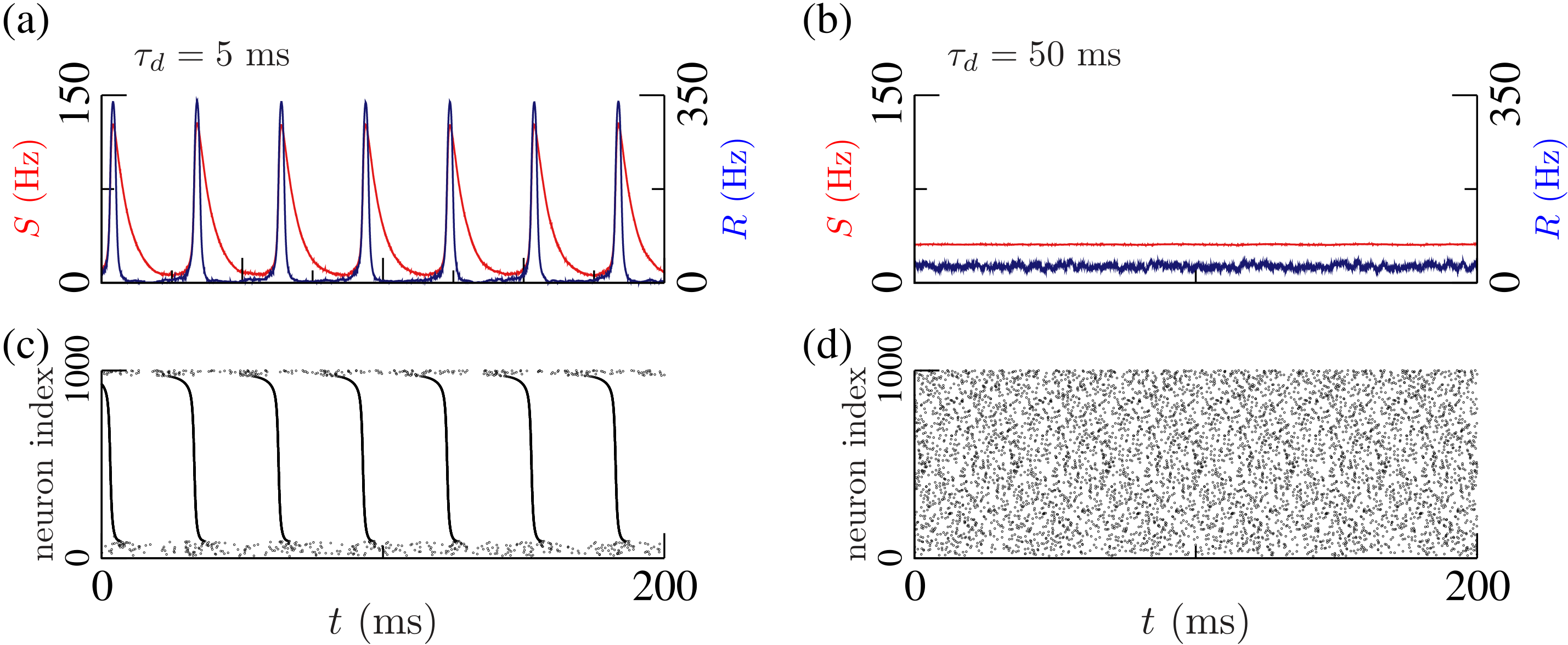}}
\caption{Networks of heterogeneous inhibitory neurons with fast synaptic kinetics 
($\tau_d=5$~ms) display macroscopic oscillations 
in the gamma range (ING oscillations)  due to collective synchronization. 
Panels (a) and (c) show the time series of the synaptic variable $S$ (red) 
and mean firing rate $R$ (blue), and the raster plot of a population of 
$N=1000$ inhibitory Wang-Buzs\'{a}ki neurons~\cite{WB96} 
with first order fast synaptic kinetics. 
The oscillations are suppressed considering  
slow inhibitory synapses ($\tau_d=50$~ms), as shown in Panels (b) and (d). 
See Materials and Methods for details on the numerical simulations.
}
\label{Fig1}
\end{figure}

Recurrent networks of spiking neurons with inhibitory interactions
readily generate fast oscillations. Figure~\ref{Fig1} shows an illustration of such
oscillations in a network of globally coupled Wang-Buzs\'aki (WB)
neurons ~\cite{WB96}. 
Panels (a,c) show the results of a  numerical simulation of the
network for fast synapses (time constant $\tau_d=5$~ms), 
compared to the membrane time constant of the neuron model ($\tau_{m} = 10$~ms).  
Although the neurons have different intrinsic frequencies due to 
a distribution in external input currents, the
raster plot reveals that fast inhibitory coupling produces the
frequency entrainment  of a large fraction of the neurons in the
ensemble. This collective  synchronization is reflected at the
macroscopic scale as an oscillation with the frequency of the
synchronous cluster of neurons~\cite{Win67,Kur84}. Indeed, panel (a) shows the time series of both the mean synaptic activation
variable  $S$, and the mean firing rate $R$, which display ING oscillations. 
Panels (b,d) of Fig.~\ref{Fig1} show the disappearance of the synchronous 
state when the synaptic kinetics is slow ($\tau_d=50$~ms).

\subsection{A heuristic firing rate equation}
A heuristic firing rate description of the spiking network simulated in 
Fig.~\ref{Fig1} takes the form~\cite{WC72,Cow14}
\begin{subequations}
\label{WC}
\begin{eqnarray}
\tau_{m} \dot R &=&-R+\Phi( -J\tau_{m}S+\Theta ),\\
\tau_{d}\dot S &=& -S+R.
\end{eqnarray}
\end{subequations}
where $R$ represents the mean firing rate in the population, $S$ is
the synaptic  activation, and the time constants $\tau_{m}$ and
$\tau_{d}$ are the neuronal and synaptic time constants 
respectively~\cite{LB11,KFR17}.
The input-output function $\Phi$, also known as the  f-I curve, is a
nonlinear function, the form of which depends on the details of the
neuronal model and on network parameters.  Finally,  $J\geq 0$ is the
synaptic strength and $\Theta $ is the mean external input current
compared to threshold. In contrast with the network model, the H-FRE
Eqs.~\eqref{WC} cannot generate sustained oscillations.  In fact, a
linear  stability analysis of steady state solutions in Eqs.~\eqref{WC}
shows that the resulting eigenvalues are 
\begin{equation}
\lambda = -\alpha (1\pm \sqrt{1-\beta} ),\label{eq:linWC}
\end{equation}
where the parameter 
$\alpha =
(\tau_{m}+\tau_{d})/(2\tau_{m}\tau_d)$ is always positive. 
Additionally, the parameter $\beta =
[4\tau_{m}\tau_{d}(1+J\tau_{m}\Phi' )]/(\tau_{m}+\tau_{d})^{2}$ 
is also positive, since the f-I curve $\Phi(x)$ is an increasing function, 
and its derivative evaluated at the steady state is then $\Phi'>0$. 
Therefore the real part of the eigenvalue 
$\lambda$ is always negative and hence steady states are always
stable, although damped oscillations are possible, 
e.g. for strong enough coupling $J$.
Introducing an explicit fixed time delay in Eqs.~\eqref{WC} can lead to
the generation of oscillations with a period on the  order of about
twice the delay~\cite{RBH05,RM11,BH08}. Nonetheless, inhibitory networks of
spiking neurons robustly show oscillations even in the absence of
explicit delays, as seen in Fig.~\ref{Fig1}. 
This suggests that there is an additional mechanism
in the network dynamics, key for driving oscillatory behavior, which 
H-FRE do not capture.

\subsection{An exact firing rate equation which captures spike synchrony}

Here we show that  the mechanism responsible for the 
generation of the oscillations shown in Fig.~\ref{Fig1} 
is the interplay between the mean firing rate and membrane potential, 
the dynamics of which reflect
the  degree of spike synchrony in the network.  To do this, we use a
set of exact macroscopic equations which have been recently derived
from a population of heterogeneous quadratic integrate-and-fire (QIF)
neurons ~\cite{MPR15}. We refer to these equations as the QIF-FRE. The
QIF-FRE with exponential  synapses have the form
\begin{subequations}
\label{qif-fre}
\begin{eqnarray}
\tau_m  \dot R &=& \frac{\Delta}{\pi\tau_m} + 2  R  V, \\ 
\tau_m \dot  V &=&  V^2 -(\pi \tau_m  R)^2  -J\tau_{m}S + \Theta,\\
\tau_{d} \dot S &=& -S+R.
\end{eqnarray}
\end{subequations}
where $\Delta$ is a parameter measuring the degree of heterogeneity in
the network and the other  parameters are as in the H-FRE Eqs.~\eqref{WC}.
Eqs.~\eqref{qif-fre} are an exact macroscopic description of the dynamics in a 
large network of heterogeneous QIF neurons with inhibitory coupling.  In contrast with the 
traditional firing rate equations Eqs.~\eqref{WC},  
they contain an explicit dependence on the subthreshold state of
the network, and hence  capture not only macroscopic variations in
firing rate, but also in spike synchrony.  Specifically, a transient
depolarizing input which drives the voltage to positive values (the
voltage has been normalized such that positive  values are
suprathrehsold, see Materials and Methods) will lead to a sharp growth
in the firing rate through the bilinear term  in Eq.~(\ref{qif-fre}a).
Simulations in the corresponding network model reveal that this
increase is due to the synchronous spiking of a subset of neurons~\cite{MPR15}.  
This increase in firing rate leads to a
hyperpolarization of the mean voltage through the quadratic term in
$R$ in Eq.~(\ref{qif-fre}b).   This term describes the effect of the
neuronal reset.  This decrease in voltage in turn drives down the mean
firing rate, and the process can repeat.  Therefore the interplay
between mean firing rate  and mean voltage in Eqs.~\eqref{qif-fre} can
generate oscillatory behavior; this behavior  corresponds to transient
bouts of spike  synchrony in the spiking network model.   

It is precisely the latency inherent in this interplay which provides
the effective  delay, which when coupled with synaptic kinetics,
generates self-sustained fast oscillations.  In fact, in the  limit of
instantaneous synapses ($\tau_{d}\to 0$), Eqs.~\eqref{qif-fre} robustly
display damped oscillations  due to the spike generation and reset
mechanism described in the preceding paragraph~\cite{MPR15}.  Contrast this  with
the dynamics in Eqs.\eqref{WC} in the same limit: the resulting H-FRE is
one dimensional and hence  cannot oscillate.

\begin{figure}[t]
\centerline{\includegraphics[width=75mm,clip=true]{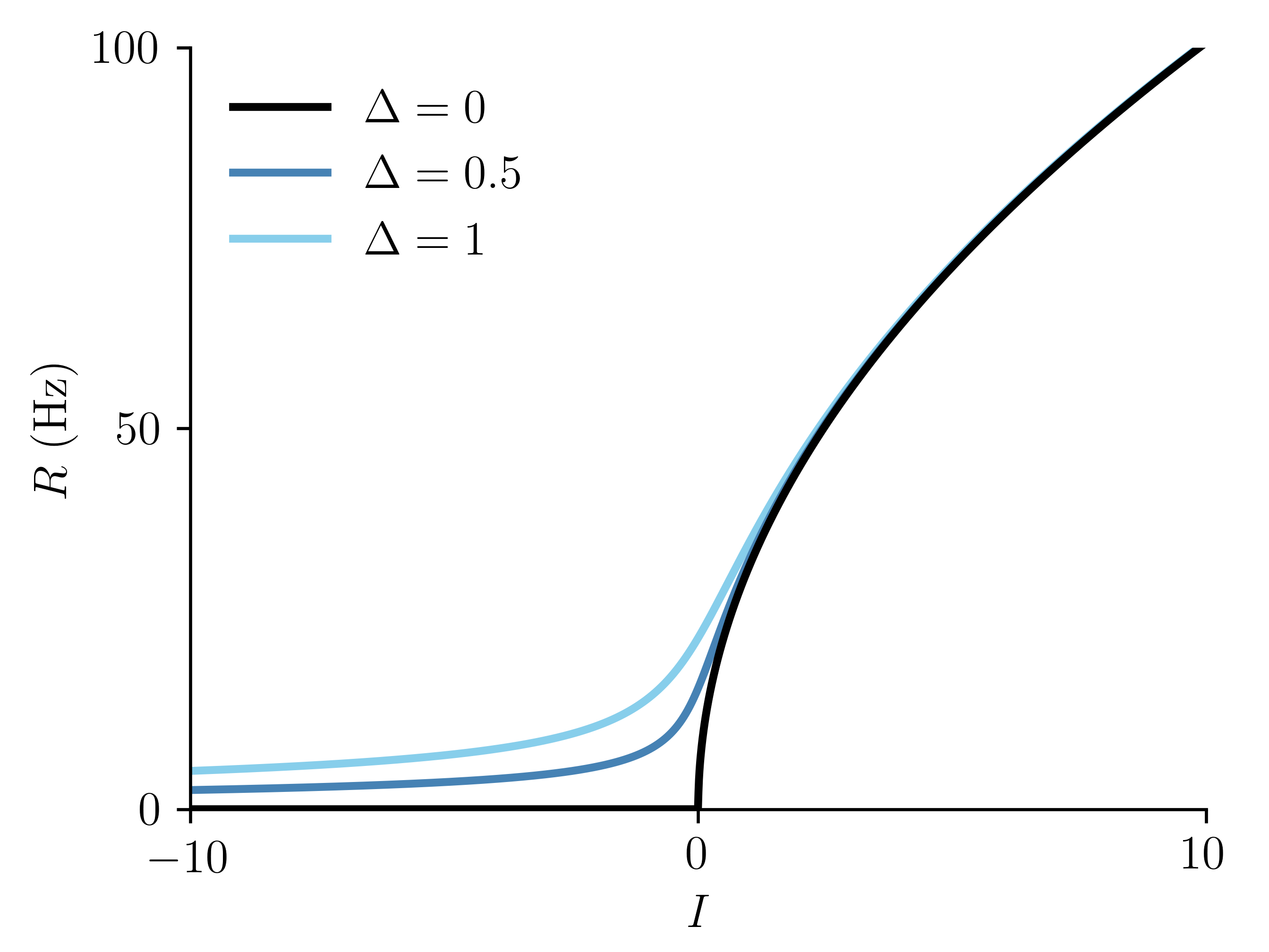}}
\caption{The f-I curve $\Phi(I)$, Eq.~\eqref{eq:fIcurve}, for several values of the 
heterogeneity parameter $\Delta $.  The membrane time constant is $\tau_m = 10$ms.}
\label{TF}
\end{figure}

On the face of things the Eqs.~\eqref{qif-fre} appear to have an utterly
distinct  functional form from the traditional Wilson-Cowan Eqs.\eqref{WC}.  
In particular, the f-I curve in H-FRE traditionally exhibits an expansive nonlinearity at low
rates and a linearization or saturation at high rates, e.g. a sigmoid.
There is no such  function visible in the QIF-FRE which have only
quadratic nonlinearities.  However, this  seeming inconsistency is
simply due to the explicit dependence of the steady-state f-I curve on
the subthreshold  voltage in Eqs.~\eqref{qif-fre}.  In fact, the
steady-state f-I curve in the QIF-FRE is ``typical'' in the
qualitative sense described above.  Specifically, solving for the steady state value of the 
firing rate in Eqs.~\eqref{qif-fre} yields 
\begin{equation}
R_*= \Phi( - J \tau_m R_*+\Theta), 
\label{Rfp}
\end{equation}
where
\begin{equation}
\Phi (I) = \frac{1}{\sqrt{2}\pi\tau_{m}}\sqrt{I+\sqrt{I^{2}+\Delta^{2}}}.\label{eq:fIcurve}
\end{equation}
The f-I curve Eq.~\eqref{eq:fIcurve} is shown in Fig.~\ref{TF} for several values of the 
parameter $\Delta$, which measures the degree of heterogeneity in the network.  
Therefore, the steady-state firing rate in Eqs.~\eqref{qif-fre}, which corresponds exactly 
to that in a network of heterogeneous QIF neurons, could easily be captured in a heuristic 
model such as Eqs.~\eqref{WC} by taking the function $\Phi $ to have the form as in 
Eq.~\eqref{eq:fIcurve}.  On the other hand, the non-steady behavior in 
Eqs.~\eqref{qif-fre}, and hence in spiking networks as well, can be quite different from 
that in the heuristic Eqs.~\eqref{WC}.

\subsubsection{Fast oscillations in the QIF-FRE}

\begin{figure}[t]
\centerline{\includegraphics[width=.9\textwidth,clip=true]{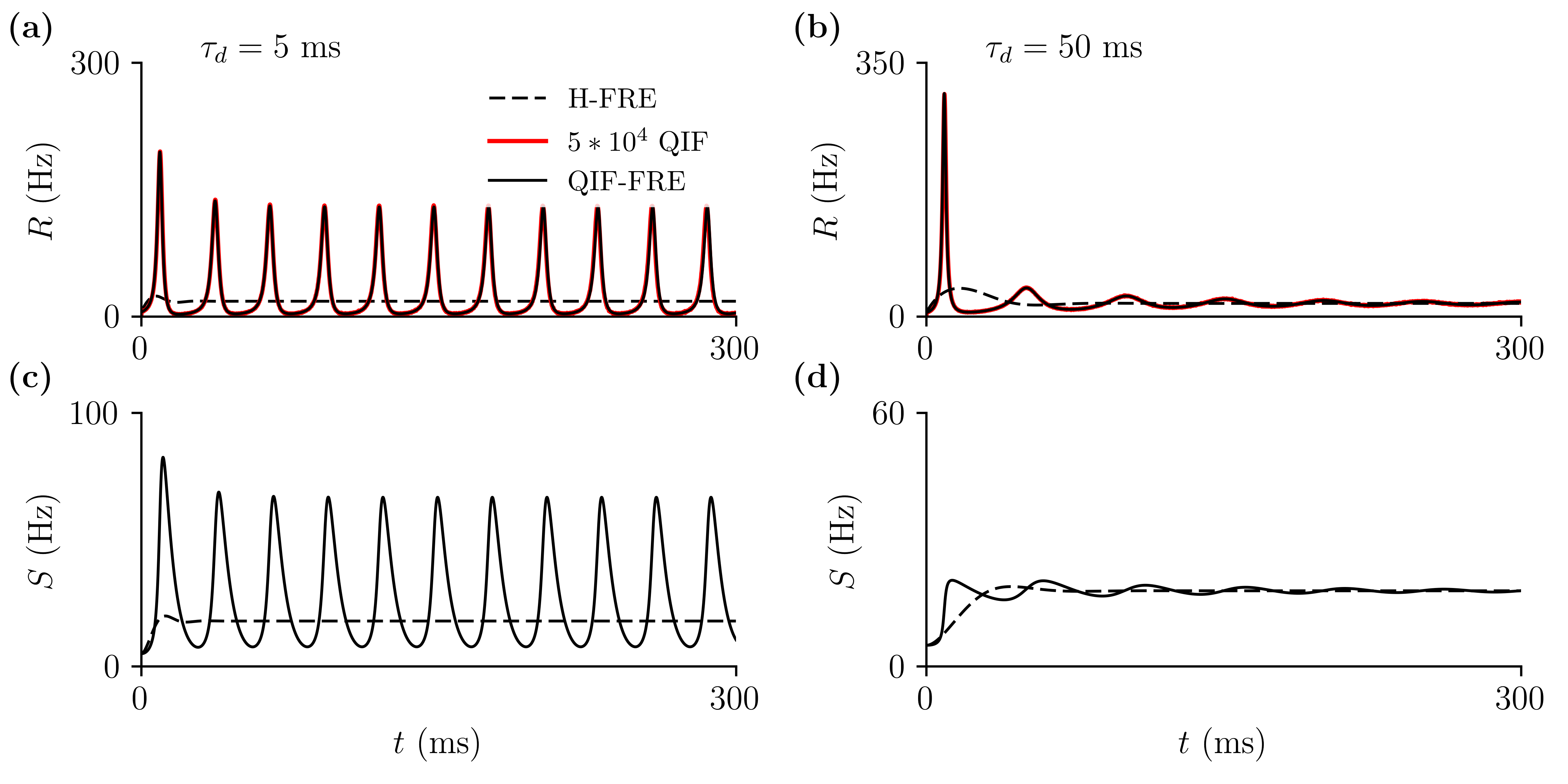}}
\caption{Heuristic FRE Eqs.~\eqref{WC} do not display inhibition-based fast 
oscillations. In contrast, networks of QIF neurons (red) and their corresponding QIF-FRE 
Eqs.~\eqref{qif-fre} (solid black) do show ING oscillations for fast synaptic 
kinetics ($\tau_d=5$~ms). These oscillations are suppressed for slow synaptic kinetics 
($\tau_d=50$~ms), as in the Wang-Buzs\'aki model shown in Fig.~\ref{Fig1}. Panels (a,b) show the times series of the Firing Rate variable $R$ of the FRE models, as well as the 
mean firing rate of a population of  $N=5\times 10 ^4$ QIF neurons (red). 
Panels (c,d) show the time series of the synaptic $S$ variables for the H-FRE (dashed line) 
and QIF-FRE (solid line). 
Parameters: $\tau_m=10$~ms, $J=21$, $\Theta=4$, $\Delta=0.3$. 
Initial values: $R(0)=S(0)=5$~Hz, $V(0)=0$.}
\label{FigSlow1}
\end{figure}

We have seen that decreasing the time constant of synaptic decay
$\tau_d$ in a network of inhibitory spiking neurons  lead to sustained
fast oscillations, while such a transition to oscillations is not
found in the heuristic rate equations, in which  the synaptic dynamics
are taken into account Eqs.~\eqref{WC}. The exact QIF-FRE, on the other
hand, do generate oscillations in this regime. Figure~\ref{FigSlow1}
shows a comparison of the firing rate $R$ and synaptic variable $S$ 
from simulations of the QIF-FRE~Eqs.\eqref{qif-fre}, with that of the 
H-FRE Eqs.~\eqref{WC}, for two different values of the  synaptic time constants. 
Additionally, we also performed simulations of a network of $N=5\times 10 ^4$ 
QIF neurons. The mean firing rate of the network is shown in red, and perfectly
agrees with the firing rate of the low dimensional QIF-FRE (solid black lines).
The function $\Phi$ in Eqs.~\eqref{WC} is chosen to be that from Eq.~\eqref{eq:fIcurve}, which is why the firing rate from both models converges to the same steady
state value when oscillations are not present (panels (b,d) 
for $\tau_d = 50$ ms).  We will study the transition to fast oscillations 
in Eqs.\eqref{qif-fre} in greater details in the following sections.

\subsection{Linear stability analysis of the QIF-FRE}

We can investigate the emergence of sustained oscillations in Eqs.~\eqref{qif-fre} 
by considering small amplitude perturbations of the steady-state solution.  
If we take $R = R_{*}+\delta Re^{\lambda t}$, 
$V = V_{*}+\delta Ve^{\lambda t}$ and $S = S_{*}+\delta Se^{\lambda t}$, 
where $\delta R$, $\delta V$, $\delta S \ll 1$, 
then the sign of the real part of the eigenvalue $\lambda $ determines whether the perturbation grows or not.  
Plugging this ansatz into Eqs.~\eqref{qif-fre} yields three coupled linear 
equations which are solvable if the following 
characteristic equation also has a solution
\begin{equation}
-2 J\tau_m R_* = (1+ \tau_{d}\lambda ) \left[(2\pi \tau_m R_*)^2 + \left(\tau_{m}\lambda  +
\frac{\Delta}{\pi\tau_{m}R_*}\right)^2 \right].
\label{ceq}
\end{equation}
The left hand side of 
Eq.~\eqref{ceq} is always negative and, for $\tau_d=0$, this implies that 
the solutions $\lambda$ are necessarily complex. 
Hence, for instantaneous synapses, the fixed point of the QIF-FRE 
is always of focus type, reflecting transient episodes 
of spike synchrony in the neuronal ensemble~\cite{MPR15}. 
In contrast, setting $\tau_d=0$ in the H-FRE, 
the system becomes first order 
and oscillations are not possible. This is the critical difference between the 
two firing rate models. 
In fact, and in contrast with the eigenvalues 
Eq.~\eqref{eq:linWC} corresponding to the growth rate of small 
perturbations in the H-FRE,
here oscillatory instabilities may occur for nonvanishing values of $\tau_d$.
Figure~\ref{PhaseD} shows the Hopf boundaries obtained from Eq.~\eqref{ceq},
as a function of the normalized synaptic strength $j  = J/\sqrt{\Theta} $ and the ratio 
of the synaptic and  neuronal time constants $\tau =\sqrt{\Theta} \tau_{d}/\tau_{m}$, 
and for different values of the ratio $\delta = \Delta/\Theta $
---see Materials and Methods, Eqs.(\ref{j}-\ref{tau}). 
The shaded regions correspond to parameter values where the QIF-FRE display oscillatory solutions.

\subsubsection{Identical neurons}
In the simplest case of identical neurons we find the boundaries of oscillatory 
instabilities explicitly. Indeed, substituting   
$\lambda=\nu+i\omega$ in Eq.~\eqref{ceq} we find that, near criticality ($ |\nu|\ll 1$), 
the real part of the eigenvalue is 
\begin{equation}
\nu \approx  J  \tau  \frac{R_*}{1+ (2 \pi \tau_d R_*)^2 }.
\label{nu}
\end{equation}
Thus, the fixed point~\eqref{Rfp}
is unstable for $ J \tau>0$, and changes its 
stability for either $J=0$, or $\tau=0$. In particular, given a non-zero 
synaptic time constant there is an oscillatory 
instability as the sign of the synaptic coupling $J$ changes from positive to negative.  
Therefore oscillations occur only for inhibitory coupling
~\cite{VAE94,Erm96,HMM95}. 
The frequency along this Hopf bifurcation line is 
determined entirely by the intrinsic firing rate of individual cells $\omega_c=2\pi R_*$. 

\begin{figure}[t]
\centerline{\includegraphics[width=120mm,clip=true]{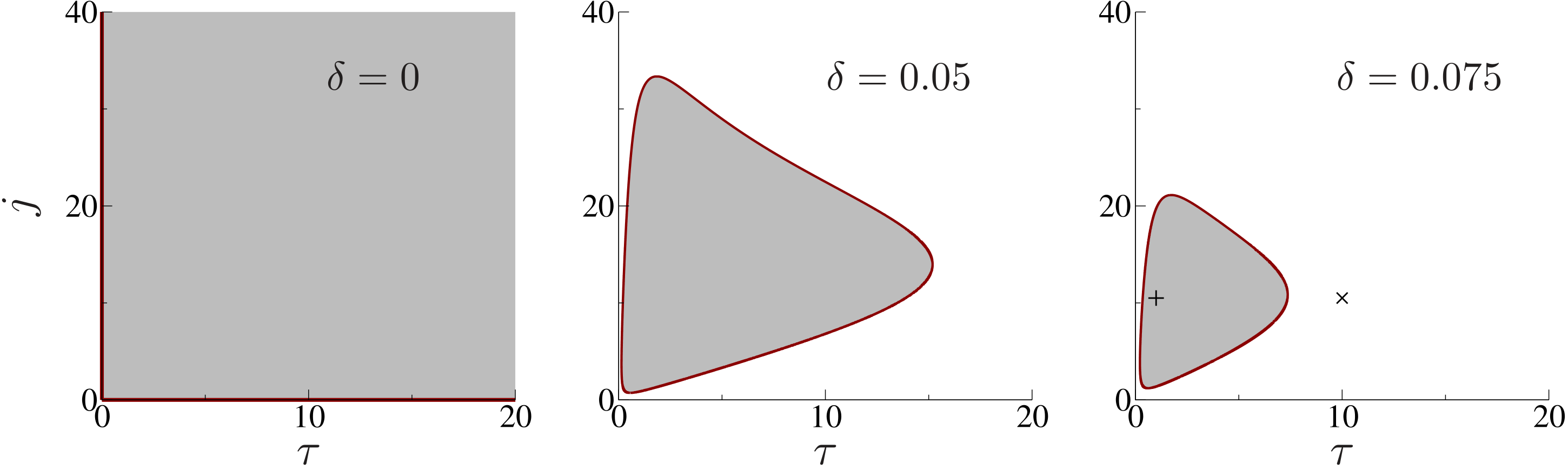}}
\caption{The ratio of the width to the center of the distribution of 
currents Eq.~\eqref{lorentzian}, $\delta=\Delta/\Theta$,
determines the presence of fast oscillations in 
the QIF-FRE. 
Oscillations disappear above the critical value given by Eq.~\eqref{deltac}. 
The panels show the Hopf boundaries of QIF-FRE with first-order synapses, 
for different values of $\delta$, obtained solving the characteristic Eq.~\eqref{ceq} 
with Re$(\lambda)=0$, see Materials and Methods. 
Shaded regions are regions of oscillations. Symbols
in the right panel correspond to the parameters used in Fig \ref{FigSlow1}.}
\label{PhaseD}
\end{figure}

On the other hand, in the limit of fast synaptic kinetics, 
i.e. $\tau_d=0$ in Eq.~\eqref{ceq}, we find another Hopf bifurcation with 
$\omega_c=\frac{1}{\tau_{m}}\sqrt{2\tau_{m}R_*(J+2\pi^2\tau_{m}R_*)}$.  This 
reflects the fact that oscillations cannot be induced if the synaptic interactions are instantaneous. 
The left panel of Figure~\ref{PhaseD} shows the phase diagram with the Hopf 
boundaries depicted in red, reflecting that oscillations are found for all values 
of inhibitory coupling and for non-instantaneous synaptic kinetics.

\subsubsection{Heterogeneous neurons}

Once heterogeneity is added to the network the region of sustained oscillatory behavior 
shrinks, see Fig.\ref{PhaseD}, center and right.  
The red closed curves correspond to the Hopf bifurcations, which have been obtained in 
parametric form from the characteristic equation~\eqref{ceq}, see Materials and Methods. 
Note that for small levels of $\delta$, oscillations are present in a 
closed region of the phase diagram, and disappear for large enough values of $\tau$ (the 
synaptic time constant relative to the neuronal time constant).  
Further increases in $\delta$ gradually reduce the region of oscillations   
until it fully disappears at the critical value 
\begin{equation}
\delta_c=\left( \frac{\Delta}{\Theta}\right)_c=\frac{1}{5}\sqrt{5-2\sqrt{5}}
=0.1453 \dots,
\label{deltac}
\end{equation}    
which has been obtained analytically from the characteristic Eq.~\eqref{ceq}, see 
Materials and Methods. 
This result is consistent with numerical studies investigating oscillations in 
heterogeneous inhibitory networks  
which indicate that 
gamma oscillations are fragile against the presence of quenched heterogeneity
~\cite{WB96,WCR+98,TJ00}.

\begin{figure}[t]
\centerline{\includegraphics[width=140mm,clip=true]{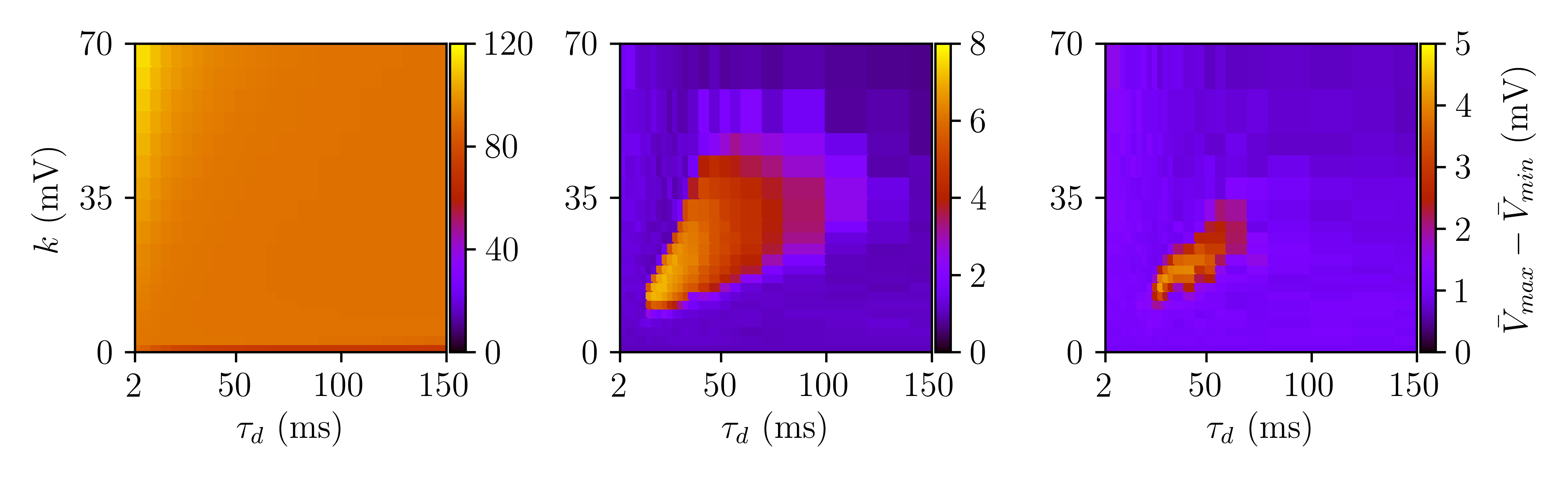}}
\vspace{0cm}
\caption{Amplitude of the oscillations of the mean membrane potential for a 
population of $N=1000$ WB neurons. From left to right: 
$\delta=\sigma/\bar I=0,0.05 ~\text{and}~0.06$. 
Central and Right panels have $\sigma=0.01 ~\mu \text{A} /\text{cm}^2$. 
See Materials and Methods for details.}
\label{WB}
\end{figure}

In the following, we compare the phase diagrams of Fig.~\ref{PhaseD} 
with numerical results using heterogeneous ensembles of Wang-Buzs\'aki neurons with 
first order synapses. 
Instead of using the population mean firing rate or mean synaptic activation,
in Fig.~\ref{WB} we computed the amplitude of the population mean 
membrane potential. This variable is less affected by finite-size fluctuations
and hence the regions of oscillations are more easily distinguishable.
The results are summarized in Fig.~\ref{WB} for different values of $\delta$ and
have been obtained by systematically increasing the coupling strength $k$
for fixed values of $\tau_d$. The resulting phase diagrams qualitatively agree 
with those shown in Fig.~\ref{PhaseD} . 
As predicted by the QIF-FRE, oscillations are found in a closed region in the ($\tau_d,k$) 
parameter space, and disappear for large enough values of $\delta$. 
Here, the critical value of $\delta=\sigma/\bar I$
is about $6$\%, smaller 
than the critical value 
given by Eq.~\eqref{deltac}. This is due to the steep 
f-I curve of the WB model, which implies a larger dispersion 
in the firing rates of the neurons even for small heterogeneities in the 
input currents. 

Additionally, for small $\tau_d$ (fast synaptic kinetics) and strong coupling $k$, 
we observed small regions where the oscillations coexist with the asynchronous 
state ---not shown. 
Numerical simulations indicate that  
this bistability is not present in the QIF-FRE. For strong coupling, and 
coexisting with the asynchronous state, we also observed various clustering states,
already reported in the original paper of Wang \& Buzs\'aki~\cite{WB96}. Clustering 
in inhibitory networks has also been observed in populations of conductance-based neurons   
with spike adaptation \cite{KE11} or time delays \cite{EPG98}. 
The fact that such states do not emerge in the model Eqs.~\eqref{qif} may be 
due to the purely sinusoidal shape of the phase resetting 
curve of the QIF model~\cite{Oku93,HMM93,KK01,Kor03,PR15,CPR16}.

\subsection{Firing Rate Equations in the limit of slow synapses}

We have seen that the oscillations which emerge in inhibitory networks 
for sufficiently fast synaptic kinetics are 
correctly described by the firing rate equations Eqs.~\eqref{qif-fre}, 
but not by the heuristic Eqs.~\eqref{WC}.  The reason for this is that the oscillations 
crucially depend on the interaction between the population firing rate and the
subthreshold membrane potential during spike initiation and reset; 
this interaction manifests itself at the network level through spike synchrony.   
Therefore, if one could suppress the spike synchrony
mechanism, and hence the  dependence on the subthreshold membrane
potential, in Eqs.~\eqref{qif-fre}, the resulting  equations ought to
bear resemblance to Eqs.~\eqref{WC}.  In fact, as we saw in Fig.~\ref{FigSlow1}, 
the two firing rate models become more similar 
when the synaptic kinetics become slower.  

Next  we  show that the two models become identical, formally, in the limit of slow
synaptic kinetics. To show this, we assume the synaptic time constant is slow, namely 
$\tau_d = \bar\tau_{d}/\epsilon $ where $0 < \epsilon\ll 1$, and rescale time 
as $\bar t = \epsilon t$.  
In this limit we are tracking the slow synaptic dynamics in 
while the neuronal dynamics are stationary to leading order,  i.e.
\begin{equation} 
R_* = \Phi (-J\tau_{m}S +\Theta ).
\label{Rslow}
\end{equation}
Therefore, the dynamics reduce to the first order equation
\begin{equation} 
\tau_d \dot S = -S+ \Phi(-J\tau_m S+\Theta).
\label{wc2}
\end{equation}
Notably, this shows that the QIF-FRE Eqs.~\eqref{qif-fre}, 
and the H-FRE~\eqref{WC}, do actually have the same dynamics in the limit of
slow synapses. In other words, Eq.~\eqref{wc2} is formally equivalent 
to the Wilson-Cowan equations for a single inhibitory population, and this 
establishes a mathematical link between the QIF-FRE and Heuristic firing rate descriptions. 
Additionally, considering slow second order synaptic kinetics (not shown),
allows one to reduce the QIF-FRE with either alpha or double exponential synapses 
to a second-order system that formally corresponds to 
the so-called neural mass models largely used for modeling EEG data, 
see e.g.~\cite{Fre75,JR95,RRW97,CGP14,ACN16}.

\begin{figure}[t]
	\centerline{\includegraphics[width=1\textwidth,clip=true]{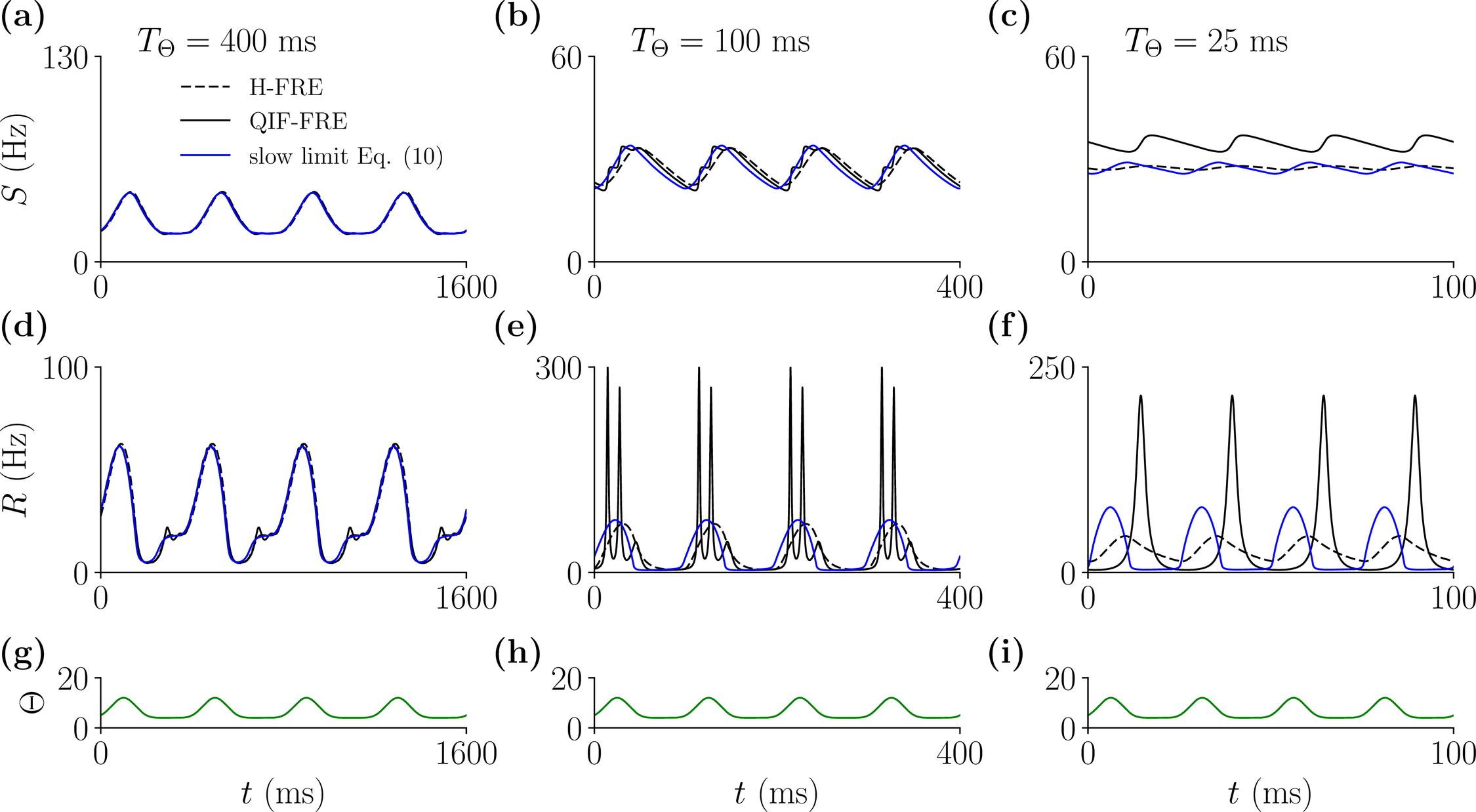}}
\caption{The reduction of the QIF-FRE to Eq.\eqref{wc2} breaks 
down when neurons receive time-varying inputs. 
Panels (a-c): $S$-variable time series for QIF-FRE (solid Black), 
H-FRE (dashed Black) and Eq.~\eqref{wc2} (Blue), 
for decreasing values of the period $T_\Theta$ of the 
external periodic forcing $\Theta(t)=4+ [1+\sin ( 2\pi  t/T_\theta )]^3$ 
---shown in panels (g-i). In all cases, the synaptic time constant is slow 
$\tau_d=100$~ms, compared to the membrane time constant $\tau_m=10$~ms.
Panels (d-f): $R$-variable time series. In the case of model Eq.~\eqref{wc2}, the
firing rate has been evaluated using Eq.\eqref{Rslow}. 
Other parameters are $J=21$, $\Delta=0.3$. 
}
\label{Fig6}
\end{figure}

\subsubsection{External inputs and breakdown of the slow-synaptic limit Eq.~\eqref{wc2}}

It is important to note that, in the derivation of Eq.~\eqref{wc2} 
we considered external inputs $\Theta$ to be constant. Then, if synapses are slow,
the neuronal variables ($R$ in the case of Eqs.~\eqref{WC} and $R$ and $V$ in the case of 
Eqs.~\eqref{qif-fre}) decay rapidly to their fixed point values.  
However even in the limit of slow synapses, such reduction can break down if external 
inputs are time-varying $\Theta=\Theta(t)$, with a time-scale which itself is not 
sufficiently slow. 

To demonstrate this, in Fig.~\ref{Fig6}, we compared the dynamics of the 
QIF-FRE and H-FRE with the approximation Eq.~\eqref{wc2}, for periodic stimuli of various 
periods ---panels (g-i)---, and always considering slow synapses, $\tau_d=100$~ms. 
As expected, the models show good agreement for very slow external inputs 
---see panels (a,d)---, but this discrepancy is strongly magnified for fast external inputs
Specifically, for fast inputs ---see panels (c,f)---, the 
dynamics of the $S$ and $R$ variables of the QIF-FRE are clearly different form those 
of the other models. 
This illustrates that even in the limit of slow synapses, the response of spiking 
networks to arbitrary time-varying inputs will always generate some degree of spike 
synchrony. 
\section{Discussion}
Firing rate models, describing the average activity of large neuronal
ensembles are broadly used in computational neuroscience. 
However, these models 
fail to  describe inhibition-based rhythms, typically observed in
networks of inhibitory neurons with synaptic kinetics
\cite{WB96,WTJ95,WCR+98,WTK+00,TJ00,BH06,BH08,BVJ07,Wan10}.
To overcome this limitation, some authors
heuristically included explicit delays in traditional FRE, and found
qualitative agreement with the oscillatory dynamics observed in
simulations of spiking neurons with both synaptic kinetics and fixed
time delays \cite{RBH05,RM11,BH08,KFR17}. Nonetheless it remains unclear why
traditional H-FRE with first order synaptic kinetics do not generate 
inhibition-based oscillations. 

Here we have investigated a novel class of FRE which can be rigorously
derived from populations of spiking (QIF) neurons~\cite{MPR15}.
Networks of globally coupled  QIF neurons with fast inhibitory
synapses readily generate fast self-sustained oscillations. 
The corresponding exact FRE, which we call the QIF-FRE, therefore also generates
oscillations. The benefit of having  a simple macroscopic description
for the network dynamics is its amenability to analysis. In
particular, the  nonlinearities in Eqs.\eqref{qif-fre}, which arise due
to the spike initiation and reset mechanism in the QIF model, conspire
to generate damped oscillations which reflect transient spike
synchrony in the network.  This oscillatory mode can be driven by
sufficiently fast recurrent inhibitory synaptic activation, leading to
sustained oscillations.  This suggests that any mean-field
description of network activity which neglects subthreshold integration 
will not  properly capture spike-synchrony-dependent dynamical
behaviors, including fast inhibitory oscillations.

The QIF-FRE have also allowed us to generate a phase diagram for
oscillatory behavior in a network of  QIF neurons with  ease via a
standard linear stability analysis, see Fig.\ref{PhaseD}. This phase
diagram agrees qualitatively with that of an equivalent network of
Wang-Buzs\'aki neurons, suggesting that the QIF-FRE not only provide
an exact description of QIF networks, but also a qualitatively
accurate description of macroscopic behaviors in networks of Class I
neurons in general. In particular, the  QIF-FRE capture the fragility
of oscillations to quenched variability in the network, 
a feature that seems to be particularly pronounced for Class 1
neuronal models compared to neural 
models with so-called Class 2 excitability~\cite{TMW+15}.

Finally we have shown that the QIF-FRE and the heuristic
H-FRE are formally equivalent in the limit of slow
synapses. In this limit the neuronal dynamics is slaved to the
synaptic activation and well-described by Eq.~\eqref{wc2},
as long as  external inputs are stationary.  
In fact, in the absence of quenched heterogeneity ($\Delta = 0$), the approximation 
for slow synapses Eq.~\eqref{wc2} is also fully equivalent to the reduction for slow 
synapses in networks of Class 1 neurons derived by Ermentrout in \cite{Erm94}.
This further indicates that the QIF-FRE are likely valid for networks
of  Class 1 neurons in general.  However, we also show that in the more
biologically plausible scenario of time-varying external drive some
degree of neuronal synchronization is generically observed, as in
Fig.~\eqref{Fig6}, and the slow-synapse reduction Eq.~\eqref{wc2} is
not valid. 

The results presented here are obtained under 
two important assumptions that need to be taken into account when comparing our work 
to the existing literature on fast oscillations in inhibitory networks. 
(\textit{i}) A derivation of an exact firing rate model for a spiking neuron network is 
only possible for ensembles of QIF neurons, which are the canonical model for 
Class 1 excitability \cite{Erm96,Izh07}. 
Although many relevant computational studies on fast inhibitory 
oscillations also consider Class 1 neurons~\cite{WB96,WCR+98,BH99,BW03,TJ00,HM03,KFR17}, 
experimental evidence indicates that fast spiking interneurons are 
highly heterogeneous in their minimal firing rates in response 
to steady currents, and that a significant fraction of them are 
Class 2~\cite{GDS+07,THR04,TR07,MLP+07} ---but see also \cite{CRT+06}. 
(\textit{ii}) Our derivation of the QIF-FRE is valid for networks of globally
coupled QIF neurons, with Lorentzian-distributed currents. 
In this system inhibition-based oscillations are only possible when the majority of the 
neurons are self-sustained oscillators, i.e. for $\Theta>0$ in 
Eq.~\eqref{lorentzian}, and are due to the 
frequency locking of a fraction of the oscillators in the population
~\cite{Win67,Kur84} ---as it can be seen in the raster plot of Fig.~\ref{Fig1}(c). 
In this state, the frequency of the cluster of synchronized oscillators coincides with 
the frequency of the mean field. The value of the frequency itself is determined 
through an interplay between single-cell resonance and network effects. Specifically, the 
synchronized neurons have intrinsic spiking frequencies near that of the mean-field oscillation 
and hence are driven resonantly.  
This collective synchronization differs from the so-called sparse synchronization 
observed in inhibitory networks of identical Class 1 neurons under the 
influence of noise~\cite{BH99,BW03,TJ00,BH08}. In the sparsely synchronized state 
neurons fire stochastically at very low rates, 
while the population firing rate displays the 
fast oscillations as the ones reported here. 

Oscillatory phenomena arising from single-cell resonances, and which reflect 
spike synchrony at the population level, are ubiquitous in 
networks of spiking neurons.  Mean-field theory for noise-driven networks leading to a Fokker-Planck formalism, allows for a linear analysis of the response of the network to weak stimuli 
when the network is in an asynchronous state~\cite{OB11,LB11}.  
Resonances can appear in the linear response when firing rates are sufficiently high or
noise strength sufficiently low. Recent work has sought to extend H-FRE in this regime by 
extracting the complex eigenvalue corresponding to the resonance and using it to construct
the linear operator of a complex-valued differential equation, the real part of which is the 
firing rate \cite{SOA13}. Other work has developed an expression for the response of
spiking networks to external drive, which often generates resonance-related damped
oscillations, through an eigenfunction expansion of the Fokker-Planck equation \cite{MD02}. 
Our approach is similar in spirit to such studies in that we also work with a low
dimensional description of the network response.  In contrast to previous work our equations
are an exact description of the macroscopic behavior, although they are only valid 
for networks of heterogeneous QIF neurons.  
Nonetheless, the QIF-FRE are simple enough to allow for an 
intuitive understanding of the origin of fast oscillations in inhibitory networks, 
and in particular, of why these oscillations are not properly captured by H-FRE.

\section{Acknowledgments}
F.D. and E.M acknowledge support 
by the European Union's Horizon 2020 research and innovation
programme under the Marie Sk{\l}odowska-Curie grant agreement No.~642563.
A.R. acknowledges a project grant from the Spanish ministry of
Economics and Competitiveness, Grant No. BFU2012-33413.
A.R. has been partially funded by the CERCA progam of the
Generalitat de Catalunya. E.M. acknowledges the projects grants 
from the Spanish ministry of
Economics and Competitiveness, Grants No.~PSI2016-75688-P and 
No.~PCIN-2015-127.


\clearpage

\section{Materials and Methods}

\subsection{Populations of inhibitory Quadratic Integrate and Fire neurons}

We model fast-spiking interneurons, the dynamics of which are
well-described by the  Hodgkin-Huxley equations with only standard
spiking currents.   
Many models of inhibitory neurons are Class 1 excitable
\cite{RE89}, including for example the  Wang-Busz\'aki (WB)
\cite{WB96}, 
and the Morris-Lecar models~\cite{ML81}. 
Class 1 models are characterized by the presence of a
saddle-node bifurcation on an invariant circle at the transition from
quiescence to spiking.  One consequence of this bifurcation structure
is the fact the spiking frequency can be arbitrarily low near
threshold. Additionally, near threshold the spiking dynamics are
dominated by the time spent in the vicinity of the saddle-node itself,
allowing for a formal reduction in dimensionality from the full
neuron model to a reduced normal form equation for a saddle-node
bifurcation~\cite{Erm96,Izh07,ET10}.  This normal form,  which is valid
for any Class 1 model near threshold, is known as the quadratic
integrate-and-fire model (QIF). 
Using a change of variables, the QIF model can be transformed to a 
phase model, called Theta-Neuron model~\cite{EK86}, which has an strictly positive 
Phase Resetting Curve (PRC). 
Neuron models with strictly positive PRC are called Type 1 neurons, 
indicating that perturbations always produce an 
advance (and not a delay) of their phase. In general, Class 1 neurons have a 
Type 1 PRC~\cite{Erm96}, but see \cite{ABC11,EGO12}.    
 
In a network of QIF neurons, the neuronal membrane potentials are
$\{\tilde V_i\}_{i=1,\ldots,N}$, which obey the  following ordinary
differential equations~\cite{EK86,LRN+00,HM03}:
\begin{equation}
C \frac{d  \tilde V_i}{dt}=g_L \frac{(\tilde V_i-V_{t})(\tilde V_i-V_{r})}{(V_t-V_r)}
+I_{0,i}
\label{qif0}
\end{equation}
where $C$ is the cell capacitance, $g_L$ is the leak conductance and $I_{0,i}$ are 
external currents. Additionally,  $V_r$ and $V_t$ represent the resting potential and threshold of the neuron, respectively. 
Using the change of variables $\tilde V'_i=\tilde V_i -(V_t+V_r)/2$, 
and then rescaling the shifted voltages as $V_i = \tilde V'_i/ (V_t-V_r)$, 
the QIF model~\eqref{qif0} reduces to 
\begin{equation}
\tau_m \dot V_i=   V_i^2 +   I_i
\label{qif} 
\end{equation}
where $\tau_m=C/g_L$ is the membrane time constant, 
$I_i=  I_{0,i}/(g_L(V_t-V_r))-1/4$ and the overdot represents derivation with 
respect to time $t$. Note that in the model~\eqref{qif} the 
voltage variables $V_i$ and the inputs $I_i$ do not have dimensions. 
Thereafter we work with QIF model its simplest form Eq.~\eqref{qif}.
We assume that the inputs are
\begin{equation}
I_i= \eta_i - J \tau_m S,
\label{input} 
\end{equation}
where $J$ is the inhibitory synaptic strength, and $S$ is the 
synaptic gating variable. Finally, the currents 
$\eta_i$ are constants taken from some prescribed distribution that here 
we consider it is a Lorentzian of half-width $\Delta$, centered at $\Theta$ 
\begin{equation}
g(\eta)= \frac{1}{\pi} \frac{\Delta}{(\eta-\Theta)^2 +\Delta^2}.
\label{lorentzian}
\end{equation}
In numerical simulations the currents were 
selected deterministically to represent the Lorentzian distribution as:
$\eta_i=\Theta+\Delta \tan(\pi/2 (2i-N-1)/(N+1))$, for $i=1,\dots,N$.
In the absence of synaptic input, the QIF model~Eqs.(\ref{qif},\ref{input})
exhibits two possible dynamical regimes, depending on the sign of $\eta_i$.
If $\eta_i<0$, the neuron is excitable, and an initial condition 
$V_i(0)<\sqrt{-\eta_i}$, asymptotically approaches  
the resting potential $-\sqrt{-\eta_i}$. For initial conditions above the 
excitability threshold, $V_i(0)>\sqrt{-\eta_i}$, the membrane potential grows without bound. 
In this case, once the neuron reaches a certain 
threshold value $V_\theta \gg 1$, it is reset to a new value $-V_\theta$ after a 
refractory period $2\tau_m/V_\theta$ (in numerical simulations, we choose $V_\theta = 100$). 
On the other hand, if $\eta_j>0$, the neuron 
behaves as an oscillator and, if $V_\theta \to \infty$,
it fires regularly with a period $T=\pi  \tau_m/\sqrt{\eta_i}$.
The instantaneous population mean firing rate is
\begin{equation}
R= \lim_{\tau_s \to 0} \frac{1}{N}\frac{1}{\tau_s} \sum_{j=1}^N  \sum_{k}
\int_{t-\tau_s}^{t}dt' \delta (t'-t_j^{k}),
\label{eq=R}
\end{equation}
where $t_j^k$ is the time of the $k$th spike of $j$th neuron, and $\delta (t)$ 
is the Dirac delta function. Finally, the dynamics of the 
synaptic variable obeys the first order ordinary differential equation 
\begin{equation}
\tau_d \dot S =-S+ R.
\label{IRsyn}
\end{equation}
For the numerical implementation of Eqs. (\ref{eq=R},\ref{IRsyn}), we set $\tau_s=10^{-2} \tau_m$. To obtain a smoother time series, the firing rate plotted in Fig. \ref{FigSlow1} was computed according to Eq. \eqref{eq=R} with $\tau_s=3 \cdot 10^{-2} \tau_m$.

\subsection{Firing Rate Equations for populations of Quadratic Integrate 
and Fire neurons}

Recently Luke et al. derived the exact macroscopic equations for a pulse-coupled ensemble of 
Theta-Neurons~\cite{LBS13}, and this has motivated a considerable number of recent papers~\cite{SLB14,Lai14,Lai15,Lai16i,Lai16ii,CB16,RM16,OS16,PD16,CHC+17}.
This work applies the so-called Ott-Antonsen theory~\cite{OA08,OA09,OHA11} to  
obtain a low-dimensional description of the network 
in terms of the complex Kuramoto order parameter. 
Nevertheless, it is is not obvious how these macroscopic descriptions 
relate to traditional H-FRE.  

As we already mentioned, the Theta-neuron model exactly 
transforms to the Quadratic Integrate and 
Fire (QIF) model by a nonlinear change of variables~\cite{EK86,Erm96,Izh07}. 
This transformation establishes a map between 
the phase variable $\theta_i \in(-\pi,\pi]$ of a Theta neuron $i$, 
and the membrane potential variable $V_i \in (-\infty,+\infty)$ of the QIF model Eq.~\eqref{qif}. 
Recently it was shown that, under some circumstances, 
a change of variables also exists at the population level~\cite{MPR15}. 
In this case, the complex Kuramoto order parameter transforms into a novel order parameter, 
composed of two macroscopic variables: 
The population-mean membrane potential $V$, and 
the population-mean firing rate $R$. As a consequence of that, 
the Ott-Antonsen theory becomes a unique method 
 for deriving exact firing rate equations  
for ensembles of heterogeneous spiking neurons ---see also~\cite{Mat16,ALB+17,SDG17} 
for recent alternative approaches.

Thus far, the FRE for QIF neurons (QIF-FRE) have been 
successfully applied to investigate the collective dynamics of 
populations of QIF neurons with instantaneous~\cite{MPR15,ERA+17,PD16},
time delayed~\cite{PM16} and excitatory synapses 
with fast synaptic kinetics~\cite{RP16}. 
However, to date the QIF-FRE have not been used to 
explore the dynamics of populations of inhibitory neurons with synaptic kinetics
---but see~\cite{CB16} for a numerical investigation using the low-dimensional 
Kuramoto order parameter description. 
The method for deriving the QIF-FRE corresponding to a population of QIF
 neurons Eq.~\eqref{qif} is exact in the 
thermodynamic limit $N\to \infty$,  
and, under the assumption that neurons are all-to-all coupled. 
Additionally, if the parameters $\eta_i$ in Eq.~\eqref{input} 
(which in the thermodynamic limit become a continuous variable) 
are assumed to be distributed according to the Lorentzian distribution Eq.~\eqref{lorentzian},
the resulting QIF-FRE become two dimensional. 
For instantaneous synapses, 
the macroscopic dynamics of the population of QIF neurons~\eqref{qif} 
is exactly described by the system of QIF-FRE~\cite{MPR15} 
\begin{subequations}
\label{fre}
\begin{eqnarray}
\tau_m  \dot R &=& \frac{\Delta}{\pi\tau_m} + 2  R  V,\label{R_eq} \\ 
\tau_m \dot  V &=&  V^2 -(\pi \tau_m  R)^2  - J \tau_m   R +\Theta , 
\label{V_eq}
\end{eqnarray}
\end{subequations}
where $R$ is the mean firing rate and $V$ the mean membrane potential in the network.  
With exponentially decaying synaptic kinetics the QIF-FRE Eqs.~\eqref{fre} become Eqs.\eqref{qif-fre}. In our study, we consider $\Theta>0$, so that the majority of the neurons are 
oscillatory ---see~Eq.\eqref{lorentzian}.

\subsubsection{Fixed points}

The fixed points of the QIF-FRE (\ref{qif-fre}) are obtained 
imposing $\dot R=\dot V=\dot S =0$. Substituting this into 
Eqs.~\eqref{qif-fre}, 
we obtain the fixed point equation $V^*=-\Delta/(2 \pi \tau_m R^*)$,
the firing rate given by Eq.~\eqref{Rfp} and $S_*=R_*$.
Note that  
for homogeneous populations, $\Delta=0$, the f-I curve Eq.~\eqref{eq:fIcurve} reduces to 
\begin{equation}
\Phi(I)=\frac{1}{\pi} \sqrt{|I|_+},
\nonumber
\end{equation}
which displays a clear threshold at $I=0$ 
(Here, $|I|_+=I$ if $I\geq0$, and vanishes for $I<0$.) This function
coincides with the \textit{squashing function} found 
by Ermentrout for homogeneous networks of Class 1 neurons~\cite{Erm94}. 
As expected, for heterogeneous networks, 
the well-defined threshold of $\Phi(I)$ for $\Delta=0$ is lost
and the transfer function becomes increasingly smoother. 

\subsubsection{Nondimensionalized QIF-FRE}
The QIF-FRE~(\ref{qif-fre}) have five parameters. It is possible to
non-dimensionalize the equations so that the system can be written solely 
in terms of 3 parameters.
Generally, we adopt the following notation: 
we use capital letters to refer to the original variables and parameters 
of the QIF-FRE, and lower case letters for non-dimensional 
quantities. A possible non-dimensionalization, valid for $\Theta>0$, is
\begin{subequations}
\begin{eqnarray}
\dot r &=& \delta/\pi + 2  r  v,  \\ 
\dot v &=&  v^2 - \pi^2 r^2  - j s +1 ,  \\
\tau \dot s &=& -s + r,
\end{eqnarray}
\end{subequations}
where the overdot here means differentiation with respect to the non-dimensional 
time $$\tilde t= \frac{\sqrt{\Theta}}{\tau_m} t.$$ 
The other variables are defined as  
\begin{equation}
r=\frac{\tau_m}{\sqrt{\Theta}} R, \quad v = \frac{V}{\sqrt{\Theta}}, \quad 
s=\frac{\tau_m}{\sqrt{\Theta}}S.
\nonumber
\end{equation}
On the other hand, the new coupling parameter is defined as 
\begin{equation}
j=\frac{J}{\sqrt{\Theta}}.
\label{j}
\end{equation}
and the parameter  
\begin{equation}
\delta= \frac{\Delta}{\Theta},
\label{delta}
\end{equation}
describes the effect of the Lorentzian 
heterogeneity \eqref{lorentzian} into the collective dynamics of 
the FRE~\eqref{fre}. Though the Lorentzian distribution does not  
have finite moments, for the sake of comparison of our results with those of 
studies investigating the dynamics of heterogeneous networks of inhibitory 
neurons, e.g.\cite{WB96,WCR+98}, the quantity $\delta$ can be compared to the 
coefficient of variation, which measures the ratio of the standard 
deviation to the mean of a probability density function. 
Finally, the non-dimensional time  
\begin{equation}
\tau= \frac{\sqrt{\Theta}}{\tau_m} \tau_{d},
\label{tau}
\end{equation}
measures the ratio of the synaptic time constant to the most-likely period 
of the neurons (times $\pi$), 
\begin{equation}
\bar T=\pi \frac{\tau_m}{\sqrt{\Theta}}.
\nonumber
\end{equation}
In numerical simulations we will use the original QIF-FRE~\eqref{qif-fre},
with $\Theta=4$, and $\tau_m=10$ms. Thus $\bar T=10 \pi/3 \approx 15.71$ms, 
so that the most likely value of the neurons' intrinsic frequency is $\bar f\approx 63.66$ Hz. 
However, our results are expressed in a more compact form in terms 
of the quantities $j,\delta,\tau$, and we will use them in some of our calculations and 
figures.

\subsection{Parametric formula for the Hopf boundaries}

To investigate the existence of oscillatory instabilities we use 
Eq.~\eqref{ceq} written in terms of the non-dimensional variables and parameters 
defined previously, which is 
\begin{equation}
-2 j  r_* = (1+ \tilde{\lambda} \tau) \left[(2\pi r_*)^2 + \left(\tilde{\lambda}  +
\frac{\delta}{\pi  r_*}\right)^2 \right].
\label{ceq2}
\end{equation}
Imposing the condition of marginal stability $\tilde{\lambda}=i\tilde{\omega}$ in 
Eq.~\eqref{ceq2} gives the system of equations
\begin{subequations}
\label{eqs}
\begin{eqnarray}
0 &=& 2 j r_*+  4 \pi^2 r_*^2 + 4 v_*^2 - (1 - 4 v_* \tau) \tilde{\omega}^2
\label{re}\\
0&=&\tilde{\omega} (4 v_* - 4 \pi^2 r_*^2 \tau - 4 v_*^2 \tau + \tau \tilde{\omega}^2)
\label{im}
\end{eqnarray}
\end{subequations}
where the fixed points are obtained from Eq.~\eqref{Rfp} solving
\begin{equation}
0=v_*^2  -\pi^2  r_*^2  -j r_*+1,
\label{fp0}
\end{equation}
with  
\begin{equation}
v_*=-\frac{\delta}{2\pi  r_*}\nonumber
\end{equation}
Eq.~\eqref{im} gives the critical frequency 
\begin{equation}
\tilde{\omega}=\frac{2}{\tau} \sqrt{ (\pi \tau r_*)^2 + \tau v_* (\tau v_* -1)  }.
\nonumber
\end{equation}
The Hopf boundaries can be plotted in parametric form 
solving Eq.~\eqref{fp0} for 
$j$, and substituting $j$ and $\tilde{\omega}$ into Eq.~\eqref{re}.
Then solving Eq.~\eqref{re} for $\tau$ gives the Hopf bifurcation boundaries  
\begin{equation}
\label{tauSols}
\tau^\pm(r_*) = \frac{ \pi^2 r_*^2 -1 + 
 7 v_*^2 \pm \sqrt{( \pi^2 r_*^2 -1)^2 - (14 + 50 \pi^2 r_*^2) v_*^2 
- 15 v_*^4} } {16 v_* (\pi^2 r_*^2 + v_*^2)}.
\end{equation}
Using the parametric formula  
\begin{equation}
\left(j(r_*),\tau^\pm(r_*) \right)^\pm= 
\left( v_*^2/r_* +1/r_* - \pi^2 r_* ,\tau^\pm (r_*)\right).
\nonumber
\end{equation}
we can be plot the Hopf boundaries 
for particular values of the parameter $\delta$, as $r_*$ is changed. 
Figure~\ref{PhaseD} shows these curves in red, for $\delta=0.05$ and $\delta=0.075$. 
They define a closed region in parameter space (shaded region) where 
oscillations are observed.

\subsubsection{Calculation of the critical value $\delta_c$, Eq~\eqref{deltac}}

The functions $\tau^\pm$ meet at two points, when the argument of the square root 
in Eq.~\eqref{tauSols} is zero. This gives four different roots for $\delta$, 
and only one of them is positive and real
\begin{equation}
\delta^* (r_*) =  \frac{2\pi r_*}{\sqrt{15}} \sqrt{ 
 8 \sqrt{1 + 5 \pi^2 r_*^2 + 10 \pi^4 r_*^4} -7 - 25 \pi^2 r_*^2}.
\nonumber
\end{equation}
This function has two positive zeros, one at $r_{*min}=0$, and one at $r_{*max}=1/\pi$,
corresponding, respectively, to the minimal ($ j \to \infty$) and maximal ($j=0$) values 
of the firing rate for identical neurons ($\delta=0$). Between these two points
the function attains a maximum, where $r_{*min}=r_{*max}=r_{*c}$, with 
$$r_{*c}=\frac{1}{ \sqrt{2\sqrt{5}}\pi }=0.1505\dots$$ 
The function $\delta^*(r_*)$ evaluated at its local maximum $r_*=r_{*c}$ gives Eq.~\eqref{deltac}.

\subsection{Populations of Wang-Buzsáki neurons}

We perform numerical simulations using the the Wang-Buzs\'{a}ki (WB) neuron \cite{WB96},
and compare them with our results using networks of QIF neurons.
The onset of oscillatory behavior in the WB model
is via a saddle node on the invariant circle (SNIC) bifurcation. 
Therefore, the populations of WB neurons near this bifurcation are expected to be well described by the theta-neuron/QIF model, the canonical model for Class 1 neural 
excitability~\cite{EK86,Erm96}. 

We numerically simulated a network of $N$ all-to-all coupled WB neurons, where the dynamics of each neuron is described by the time evolution of its membrane potential~\cite{WB96}
\begin{equation*}
C_m\dot{V_i}=-I_{\text{Na},i}-I_{\text{K},i}-I_{\text{L},i}-I_{\text{syn}}+I_{\text{app},i}+I_{0}.
\end{equation*}
The cell capacitance is $C_m=1~\mu \text{F}/\text{cm}^2$. 
The inputs $I_\text{app}$ (in $\mu \text{A} /\text{cm}^2$) are distributed according to 
a Lorentzian distribution with half width $\sigma$ and center $\bar{I}$.
In numerical simulations these currents were 
selected deterministically to represent the Lorentzian distribution as
$I_{\text{app},i}=\bar I+\sigma \tan(\pi/2 (2i-N-1)/(N+1))$, for $i=1,\dots,N$.
The constant input $I_{0}=0.1601 ~\mu \text{A} /\text{cm}^2$ sets the neuron at the SNIC bifurcation when $I_{app}=0$. The leak current is
\begin{equation*}
I_{\text{L},i}=g_\text{L}\left(V_i-E_\text{L}\right),
\end{equation*}
with $g_\text{L}=0.1 \text{ mS}/ \text{cm}^2$, so that the passive time constant $\tau_m=C_m/g_\text{L}=10 \text{ ms}$. 
The sodium current is 
\begin{equation*}
 I_{\text{Na},i}=g_{\text{Na}} m_{\infty}^{3} h \left(V_i-E_{\text{Na}} \right),
\end{equation*}
where $g_{\text{Na}}=35 \text{ mS}/{\text{cm}^2}$, $E_{\text{Na}}=55 \text{ mV}$,  $m_{\infty}= \alpha_{m}/\left(\alpha_{m}+\beta_{m}\right)$ with $\alpha_{m}\left(V_{i}\right)=-0.1\left(V_i+\allowbreak 35\right)/\left(\exp{\left(-0.1\left(V_i+\allowbreak 35\right)-1\right)}\right)$, $\beta_{m}\left(V_{i}\right){=}4\exp\left({-}\left(V_{i}{+}60\right){/}18\right)$. 
The inactivation variable $h$ obeys the differential equation
\begin{equation*}
\dot{h}=\phi\left(\alpha_h\left(1-h\right)-\beta_hh\right),
\end{equation*}
with $\phi=5$, $\alpha_h\left(V_{i}\right)=0.07\exp\left(-\left(V_{i}+58\right)/20\right)$ and $\beta_h\left(V_{i}\right)=1/\left(\exp\left(-0.1\left(V_{i}+ 28\right)\right)+1\right)$.
The potassium current follows
\begin{equation*}
I_{\text{K},i}=g_{\text{K}}n^4\left(V_{i}-E_{\text{K}}\right),
\end{equation*}
with $g_\text{K}=9 \text{ mS}/\text{cm}^2$, $E_{\text{K}}=-90 \text{ mV}$. The activation variable $n$ obeys
\begin{equation*}
\dot{n}=\phi\left(\alpha_n\left(1-n\right)-\beta_nn\right),
\end{equation*}
where $\alpha_n\left(V_{i}\right)=-0.01\left(V_i+34\right)/\left(\exp\left(-0.1\left(V_i+34\right)\right)-1\right)$ and $\beta_n\left(V_i\right)=0.125\exp\left(-\left(V_i+44\right)/80\right)$.

The synaptic current is $I_\text{syn}=k C_mS$, where the synaptic activation variable $S$ obeys the first order kinetics Eq.~\eqref{IRsyn} and $k$ is the coupling strength (expressed in mV). The factor $C_m$ ensures 
that the effect of an incoming spike to the neuron is independent from its passive time constant. The neuron is defined to emit a spike when its membrane potential crosses $0$ mV. The population firing rate is then computed according to Eq. \eqref{eq=R}, with $\tau_s=10^{-2}~ \text{ms}$. 
In numerical simulations we considered $N=1000$ all-to-all coupled WB neurons, 
using the Euler method with time step $dt=0.001~\text{ms}$.
In Fig. \ref{Fig1}, the membrane potentials were initially randomly distributed according to a Lorentzian function
 with half width $5 ~ \text{mV}$ and center $-62 ~\text{mV}$. 
Close to the bifurcation point, this is equivalent to uniformly distribute the phases of the corresponding Theta-Neurons in $\left[-\pi,\pi\right]$\cite{EK86,Izh07,ET10,LRN+00}. The parameters were chosen as $\bar{I}=0.5 ~\mu\text{A}/\text{cm}^2$, $\sigma=0.01~\mu\text{A}/\text{cm}^2$ and $k=6 ~\text{mV} $. The population firing rate was smoothed setting $\tau_s=2$ ms in Eq.\eqref{eq=R}. \\
In Fig. \ref{WB}, we systematically varied the coupling strength and the synaptic time decay constant to determine the range of parameters displaying oscillatory behavior. 
For each fixed value of $\tau_d$ we varied the coupling strength $k$; we performed two series of simulations, for increasing and decreasing coupling strength. 
In Fig.~\ref{WB} we only show results for increasing $k$. 

All quantities were measured after a transient of $1000~\text{ms}$. 
To obtain the amplitude of the oscillations of the mean membrane potential, 
we computed the maximal amplitude $\bar{V}_{\text{max}}-\bar{V}_{\text{min}}$ 
over time windows of $200~\text{ms}$ for $1000~\text{ms}$, 
and then averaged over the five windows.

\clearpage


\begin{thebibliography}{10}



\bibitem{WC72}
Wilson HR, Cowan JD.
\newblock Excitatory and inhibitory interactions in localized populations of
  model neurons.
\newblock Biophys J. 1972;12(1):1--24.

\bibitem{ET10}
Ermentrout GB, Terman DH.
\newblock Mathematical foundations of neuroscience. vol.~64.
\newblock Springer; 2010.

\bibitem{GKN+14}
Gerstner W, Kistler WM, Naud R, Paninski L.
\newblock Neuronal dynamics: From single neurons to networks and models of
  cognition.
\newblock Cambridge University Press; 2014.

\bibitem{DA01}
Dayan P, Abbott LF.
\newblock Theoretical neuroscience.
\newblock Cambridge, MA: MIT Press; 2001.

\bibitem{Cow14}
Cowan J.
\newblock A personal account of the development of the field theory of
  large-scale brain activity from 1945 onward.
\newblock In: Neural fields. Springer; 2014. p. 47--96.

\bibitem{CGP14}
Coombes S, beim Graben P, Potthast R.
\newblock Tutorial on neural field theory.
\newblock In: Neural fields. Springer; 2014. p. 1--43.

\bibitem{LRN+00}
Latham P, Richmond B, Nelson P, Nirenberg S.
\newblock Intrinsic dynamics in neuronal networks. I. Theory.
\newblock Journal of Neurophysiology. 2000;83(2):808--827.

\bibitem{SHS03}
Shriki O, Hansel D, Sompolinsky H.
\newblock Rate models for conductance-based cortical neuronal networks.
\newblock Neural Comput. 2003;15(8):1809--1841.

\bibitem{RBH05}
Roxin A, Brunel N, Hansel D.
\newblock Role of delays in shaping spatiotemporal dynamics of neuronal
  activity in large networks.
\newblock Phys Rev Lett. 2005;94(23):238103.

\bibitem{RM11}
Roxin A, Montbri{\'o} E.
\newblock How effective delays shape oscillatory dynamics in neuronal networks.
\newblock Physica D. 2011;240(3):323--345.

\bibitem{WC73}
Wilson HR, Cowan JD.
\newblock A mathematical theory of the functional dynamics of cortical and
  thalamic nervous tissue.
\newblock Kybernetik. 1973;13(2):55--80.
\newblock doi:{10.1007/BF00288786}.

\bibitem{Ama74}
Amari Si.
\newblock A method of statistical neurodynamics.
\newblock Kybernetik. 1974;14(4):201--215.
\newblock doi:{10.1007/BF00274806}.

\bibitem{Nun74}
Nunez PL.
\newblock The brain wave equation: a model for the EEG.
\newblock Mathematical Biosciences. 1974;21(3):279 -- 297.
\newblock doi:{http://dx.doi.org/10.1016/0025-5564(74)90020-0}.

\bibitem{EC79}
Ermentrout GB, Cowan JD.
\newblock A mathematical theory of visual hallucination patterns.
\newblock Biological Cybernetics. 1979;34(3):137--150.
\newblock doi:{10.1007/BF00336965}.

\bibitem{BLS95}
Ben-Yishai R, Bar-Or RL, Sompolinsky H.
\newblock Theory of orientation tuning in visual cortex.
\newblock Proc Nat Acad Sci. 1995;92(9):3844--3848.

\bibitem{PBS+96}
Pinto DJ, Brumberg JC, Simons DJ, Ermentrout GB, Traub R.
\newblock A quantitative population model of whisker barrels: Re-examining the
  Wilson-Cowan equations.
\newblock Journal of Computational Neuroscience. 1996;3(3):247--264.
\newblock doi:{10.1007/BF00161134}.

\bibitem{HS98}
Hansel D, Sompolinsky H.
\newblock Modeling Feature Selectivity in Local Cortical Circuits.
\newblock In: Koch C, Segev I, editors. Methods in Neuronal Modelling: From
  Ions to Networks. Cambridge: MIT Press; 1998. p. 499--567.

\bibitem{TPM98}
Tsodyks~M MH Pawelzik~K.
\newblock Neural networks with dynamic synapses.
\newblock Neural Comput. 1998;10:821.

\bibitem{Wil99}
Wilson HR.
\newblock Spikes, decisions, and actions: the dynamical foundations of
  neurosciences. 1999;.

\bibitem{TSO+00}
Tabak J, Senn W, O’Donovan MJ, Rinzel J.
\newblock Modeling of spontaneous activity in developing spinal cord using
  activity-dependent depression in an excitatory network.
\newblock J Neurosci. 2000;20:3041--3056.

\bibitem{BCG+01}
Bressloff PC, Cowan JD, Golubitsky M, Thomas PJ, Wiener MC.
\newblock Geometric visual hallucinations, Euclidean symmetry and the
  functional architecture of striate cortex.
\newblock Philosophical Transactions of the Royal Society of London B:
  Biological Sciences. 2001;356(1407):299--330.
\newblock doi:{10.1098/rstb.2000.0769}.

\bibitem{LTG+02}
Laing CR, Troy WC, Gutkin B, Ermentrout GB.
\newblock Multiple bumps in a neuronal model of working memory.
\newblock SIAM Journal on Applied Mathematics. 2002;63(1):62--97.

\bibitem{HT06}
Holcman D, Tsodyks M.
\newblock The emergence of up and down states in cortical networks.
\newblock PLoS Comput Biol. 2006;2(3):e23.

\bibitem{MRR07}
Moreno-Bote R, Rinzel J, Rubin N.
\newblock Noise-induced alternations in an attractor network model of
  perceptual bistability.
\newblock J Neurophysiol. 2007;98(3):1125--1139.

\bibitem{MBT08}
Mongillo G, Barak O, Tsodyks M.
\newblock Synaptic theory of working memory.
\newblock Science. 2008;319(5869):1543--1546.

\bibitem{TWC+11}
Touboul J, Wendling F, Chauvel P, Faugeras O.
\newblock Neural mass activity, bifurcations, and epilepsy.
\newblock Neural computation. 2011;23(12):3232--3286.

\bibitem{MR13}
Mart{\'\i} D, Rinzel J.
\newblock Dynamics of feature categorization.
\newblock Neural computation. 2013;25(1):1--45.

\bibitem{TDD14}
Ton R, Deco G, Daffertshofer A.
\newblock Structure-function discrepancy: inhomogeneity and delays in
  synchronized neural networks.
\newblock PLOS Comput Biol. 2014;10(7):e1003736.

\bibitem{SOA13}
Schaffer ES, Ostojic S, Abbott L.
\newblock A Complex-Valued Firing-Rate Model That Approximates the Dynamics of
  Spiking Networks.
\newblock PLoS Comput Biol. 2013;9(10):e1003301.

\bibitem{WB96}
Wang XJ, Buzs{\'a}ki G.
\newblock Gamma oscillation by synaptic inhibition in a hippocampal
  interneuronal network model.
\newblock The journal of Neuroscience. 1996;16(20):6402--6413.

\bibitem{WTJ95}
Whittington MA, Traub RD, Jefferys JG.
\newblock Synchronized oscillations in interneuron networks driven by
  metabotropic glutamate receptor activation.
\newblock Nature. 1995;373:612--615.

\bibitem{WCR+98}
White JA, Chow CC, Rit J, Soto-Trevi{\~n}o C, Kopell N.
\newblock Synchronization and oscillatory dynamics in heterogeneous, mutually
  inhibited neurons.
\newblock Journal of computational neuroscience. 1998;5(1):5--16.

\bibitem{WTK+00}
Whittington MA, Traub RD, Kopell N, Ermentrout B, Buhl EH.
\newblock Inhibition-based rhythms: experimental and mathematical observations
  on network dynamics.
\newblock Int Journal of Psychophysiol. 2000;38(3):315 -- 336.
\newblock doi:{http://dx.doi.org/10.1016/S0167-8760(00)00173-2}.

\bibitem{TJ00}
Tiesinga P, Jos{\'e} JV.
\newblock Robust gamma oscillations in networks of inhibitory hippocampal
  interneurons.
\newblock Network: Computation in Neural Systems. 2000;11(1):1--23.

\bibitem{BH06}
Brunel N, Hansel D.
\newblock How noise affects the synchronization properties of recurrent
  networks of inhibitory neurons.
\newblock Neural Comput. 2006;18(5):1066--1110.

\bibitem{BH08}
Brunel N, Hakim V.
\newblock Sparsely synchronized neuronal oscillations.
\newblock Chaos: An Interdisciplinary Journal of Nonlinear Science.
  2008;18(1):015113.

\bibitem{BVJ07}
Bartos M, Vida I, Jonas P.
\newblock Synaptic mechanisms of synchronized gamma oscillations in inhibitory
  interneuron networks.
\newblock Nature reviews neuroscience. 2007;8(1):45--56.

\bibitem{Wan10}
Wang XJ.
\newblock Neurophysiological and computational principles of cortical rhythms
  in cognition.
\newblock Physiological reviews. 2010;90(3):1195--1268.

\bibitem{KFR17}
Keeley S, Fenton AA, Rinzel J.
\newblock Modeling fast and slow gamma oscillations with interneurons of
  different subtype.
\newblock Journal of Neurophysiology. 2017;117(3):950--965.
\newblock doi:{10.1152/jn.00490.2016}.

\bibitem{MPR15}
Montbri\'o E, Paz\'o D, Roxin A.
\newblock Macroscopic Description for Networks of Spiking Neurons.
\newblock Phys Rev X. 2015;5:021028.
\newblock doi:{10.1103/PhysRevX.5.021028}.

\bibitem{Win67}
Winfree AT.
\newblock Biological rhythms and the behavior of populations of coupled
  oscillators.
\newblock J Theor Biol. 1967;16:15--42.

\bibitem{Kur84}
Kuramoto Y.
\newblock Chemical Oscillations, Waves, and Turbulence.
\newblock Berlin: {S}pringer-{V}erlag; 1984.

\bibitem{LB11}
Ledoux E, Brunel N.
\newblock Dynamics of networks of excitatory and inhibitory neurons in response
  to time-dependent inputs.
\newblock Frontiers Comp Neurosci. 2011;5:25.

\bibitem{VAE94}
Van~Vreeswijk C, Abbott LF, Bard~Ermentrout G.
\newblock When inhibition not excitation synchronizes neural firing.
\newblock Journal of Computational Neuroscience. 1994;1(4):313--321.
\newblock doi:{10.1007/BF00961879}.

\bibitem{Erm96}
Ermentrout B.
\newblock Type I membranes, phase resetting curves, and synchrony.
\newblock Neural Comp. 1996;8:979--1001.

\bibitem{HMM95}
Hansel D, Mato G, Meunier C.
\newblock Synchrony in excitatory neural networks.
\newblock Neural Comput. 1995;7:307--337.

\bibitem{KE11}
Kilpatrick ZP, Ermentrout B.
\newblock Sparse Gamma Rhythms Arising through Clustering in Adapting Neuronal
  Networks.
\newblock PLoS Comput Biol. 2011;7(11):e1002281.
\newblock doi:{10.1371/journal.pcbi.1002281}.

\bibitem{EPG98}
Ernst U, Pawelzik K, Geisel T.
\newblock Delay-induced multistable synchronization of biological oscillators.
\newblock Physical review E. 1998;57(2):2150.

\bibitem{Oku93}
Okuda K.
\newblock Variety and generality of clustering in globally coupled oscillators.
\newblock Physica D: Nonlinear Phenomena. 1993;63(3-4):424--436.

\bibitem{HMM93}
Hansel D, Mato G, Meunier C.
\newblock Clustering and slow switching in globally coupled phase oscillators.
\newblock Phys Rev E. 1993;48:3470--3477.
\newblock doi:{10.1103/PhysRevE.48.3470}.

\bibitem{KK01}
Kori H, Kuramoto Y.
\newblock Slow switching in globally coupled oscillators: robustness and
  occurrence through delayed coupling.
\newblock Phys Rev E. 2001;63:046214.
\newblock doi:{10.1103/PhysRevE.63.046214}.

\bibitem{Kor03}
Kori H.
\newblock Slow switching and broken cluster state in a population of neuronal
  oscillators.
\newblock Int J Mod Phys B. 2003;17:4238--4241.
\newblock doi:{10.1142/S0217979203022246}.

\bibitem{PR15}
Politi A, Rosenblum M.
\newblock Equivalence of phase-oscillator and integrate-and-fire models.
\newblock Phys Rev E. 2015;91:042916.
\newblock doi:{10.1103/PhysRevE.91.042916}.

\bibitem{CPR16}
Clusella P, Politi A, Rosenblum M.
\newblock A minimal model of self-consistent partial synchrony.
\newblock New J Phys. 2016;18(9):093037.

\bibitem{Fre75}
Freeman WJ.
\newblock Mass action in the nervous system.
\newblock Academic Press, New York; 1975.

\bibitem{JR95}
Jansen BH, Rit VG.
\newblock Electroencephalogram and visual evoked potential generation in a
  mathematical model of coupled cortical columns.
\newblock Biological Cybernetics. 1995;73(4):357--366.
\newblock doi:{10.1007/BF00199471}.

\bibitem{RRW97}
Robinson PA, Rennie CJ, Wright JJ.
\newblock Propagation and stability of waves of electrical activity in the
  cerebral cortex.
\newblock Phys Rev E. 1997;56:826--840.
\newblock doi:{10.1103/PhysRevE.56.826}.

\bibitem{ACN16}
Ashwin P, Coombes S, Nicks R.
\newblock Mathematical Frameworks for Oscillatory Network Dynamics in
  Neuroscience.
\newblock The Journal of Mathematical Neuroscience. 2016;6(1):1--92.
\newblock doi:{10.1186/s13408-015-0033-6}.

\bibitem{TMW+15}
Tikidji-Hamburyan RA, Mart{\'\i}nez JJ, White JA, Canavier CC.
\newblock Resonant Interneurons Can Increase Robustness of Gamma Oscillations.
\newblock Journal of Neuroscience. 2015;35(47):15682--15695.
\newblock doi:{10.1523/JNEUROSCI.2601-15.2015}.

\bibitem{Erm94}
Ermentrout B.
\newblock Reduction of conductance-based models with slow synapses to neural
  nets.
\newblock Neural Comput. 1994;6(4):679--695.

\bibitem{Izh07}
Izhikevich EM.
\newblock Dynamical Systems in Neuroscience.
\newblock Cambridge, Massachusetts: The MIT Press; 2007.

\bibitem{BH99}
Brunel N, Hakim V.
\newblock Fast global oscillations in networks of integrate-and-fire neurons
  with low firing rates.
\newblock Neural Comput. 1999;11(7):1621--1671.

\bibitem{BW03}
Brunel N, Wang XJ.
\newblock What determines the frequency of fast network oscillations with
  irregular neural discharges? I. Synaptic dynamics and excitation-inhibition
  balance.
\newblock Journal of neurophysiology. 2003;90(1):415--430.

\bibitem{HM03}
Hansel D, Mato G.
\newblock Asynchronous states and the emergence of synchrony in large networks
  of interacting excitatory and inhibitory neurons.
\newblock Neural Computation. 2003;15(1):1--56.

\bibitem{GDS+07}
Golomb D, Donner K, Shacham L, Shlosberg D, Amitai Y, Hansel D.
\newblock Mechanisms of firing patterns in fast-spiking cortical interneurons.
\newblock PLoS Computational Biology. 2007;3(8):e156.

\bibitem{THR04}
Tateno T, Harsch A, Robinson HPC.
\newblock Threshold Firing Frequency{\textendash}Current Relationships of
  Neurons in Rat Somatosensory Cortex: Type 1 and Type 2 Dynamics.
\newblock Journal of Neurophysiology. 2004;92(4):2283--2294.
\newblock doi:{10.1152/jn.00109.2004}.

\bibitem{TR07}
Tateno T, Robinson HPC.
\newblock Phase Resetting Curves and Oscillatory Stability in Interneurons of
  Rat Somatosensory Cortex.
\newblock Biophys J. 2007;92(2):683--695.
\newblock doi:{10.1529/biophysj.106.088021}.

\bibitem{MLP+07}
Mancilla JG, Lewis TJ, Pinto DJ, Rinzel J, Connors BW.
\newblock Synchronization of Electrically Coupled Pairs of Inhibitory
  Interneurons in Neocortex.
\newblock Journal of Neuroscience. 2007;27(8):2058--2073.
\newblock doi:{10.1523/JNEUROSCI.2715-06.2007}.

\bibitem{CRT+06}
La~Camera G, Rauch A, Thurbon D, L{\"u}scher HR, Senn W, Fusi S.
\newblock Multiple Time Scales of Temporal Response in Pyramidal and Fast
  Spiking Cortical Neurons.
\newblock Journal of Neurophysiology. 2006;96(6):3448--3464.
\newblock doi:{10.1152/jn.00453.2006}.

\bibitem{OB11}
Ostojic S, Brunel N.
\newblock From spiking neuron models to linear-nonlinear models.
\newblock PLoS Comput Biol. 2011;7(1):e1001056.

\bibitem{MD02}
Mattia M, Del~Giudice P.
\newblock Population dynamics of interacting spiking neurons.
\newblock Phys Rev E. 2002;66:051917.
\newblock doi:{10.1103/PhysRevE.66.051917}.

\bibitem{RE89}
Rinzel J, Ermentrout B.
\newblock Analysis of neural excitability and oscillations.
\newblock In: Koch C, Segev I, editors. Methods in Neuronal Modelling: From
  Ions to Networks. Cambridge: MIT Press; 1989. p. 135--171.

\bibitem{ML81}
Morris C, Lecar H.
\newblock Voltage oscillations in the barnacle giant muscle fiber.
\newblock Biophysical journal. 1981;35(1):193--213.

\bibitem{EK86}
Ermentrout B, Kopell N.
\newblock Parabolic bursting in an excitable system coupled with a slow
  oscillation.
\newblock {SIAM} J Appl Math. 1986;46:233--253.

\bibitem{ABC11}
Achuthan S, Butera RJ, Canavier CC.
\newblock Synaptic and intrinsic determinants of the phase resetting curve for
  weak coupling.
\newblock Journal of Computational Neuroscience. 2011;30(2):373--390.
\newblock doi:{10.1007/s10827-010-0264-1}.

\bibitem{EGO12}
Ermentrout GB, Glass L, Oldeman BE.
\newblock The Shape of Phase-Resetting Curves in Oscillators with a Saddle Node
  on an Invariant Circle Bifurcation.
\newblock Neural Computation. 2012;24(12):3111--3125.
\newblock doi:{10.1162/NECO\_a\_00370}.

\bibitem{LBS13}
Luke TB, Barreto E, So P.
\newblock Complete classification of the macroscopic behavior of a
  heterogeneous network of theta neurons.
\newblock Neural Comput. 2013;25(12):3207--3234.

\bibitem{SLB14}
So P, Luke TB, Barreto E.
\newblock Networks of theta neurons with time-varying excitability: Macroscopic
  chaos, multistability, and final-state uncertainty.
\newblock Physica D. 2014;267(0):16--26.
\newblock doi:{http://dx.doi.org/10.1016/j.physd.2013.04.009}.

\bibitem{Lai14}
Laing CR.
\newblock Derivation of a neural field model from a network of theta neurons.
\newblock Phys Rev E. 2014;90:010901.
\newblock doi:{10.1103/PhysRevE.90.010901}.

\bibitem{Lai15}
Laing CR.
\newblock Exact Neural Fields Incorporating Gap Junctions.
\newblock SIAM Journal on Applied Dynamical Systems. 2015;14(4):1899--1929.

\bibitem{Lai16i}
Laing CR.
\newblock Travelling waves in arrays of delay-coupled phase oscillators.
\newblock Chaos. 2016;26(9).
\newblock doi:{http://dx.doi.org/10.1063/1.4953663}.

\bibitem{Lai16ii}
Laing CR.
\newblock Bumps in Small-World Networks.
\newblock Frontiers in Computational Neuroscience. 2016;10:53.
\newblock doi:{10.3389/fncom.2016.00053}.

\bibitem{CB16}
Coombes S, Byrne {\'A}.
\newblock Next generation neural mass models.
\newblock in Lecture Notes in Nonlinear Dynamics in Computational Neuroscience: from Physics and Biology to ICT Springer (In Press).

\bibitem{RM16}
Roulet J, Mindlin GB.
\newblock Average activity of excitatory and inhibitory neural populations.
\newblock Chaos: An Interdisciplinary Journal of Nonlinear Science.
  2016;26(9):093104.
\newblock doi:{10.1063/1.4962326}.

\bibitem{OS16}
O'Keeffe KP, Strogatz SH.
\newblock Dynamics of a population of oscillatory and excitable elements.
\newblock Phys Rev E. 2016;93:062203.
\newblock doi:{10.1103/PhysRevE.93.062203}.

\bibitem{PD16}
Pietras B, Daffertshofer A.
\newblock Ott-Antonsen attractiveness for parameter-dependent oscillatory
  systems.
\newblock Chaos: An Interdisciplinary Journal of Nonlinear Science.
  2016;26(10):103101.
\newblock doi:{10.1063/1.4963371}.

\bibitem{ERA+17}
Esnaola-Acebes JM, Roxin A,  Avitabile D, Montbri\'o E.
\newblock Synchrony-induced modes of oscillation of a neural field model.
\newblock Phys Rev E. 2017;96:052407.
\newblock doi:{10.1103/PhysRevE.96.052407}.

\bibitem{CHC+17}
Chandra S, Hathcock D, Crain K, Antonsen TM, Girvan M, Ott E.
\newblock Modeling the network dynamics of pulse-coupled neurons.
\newblock Chaos: An Interdisciplinary Journal of Nonlinear Science.
  2017;27(3):033102.
\newblock doi:{10.1063/1.4977514}.

\bibitem{OA08}
Ott E, Antonsen TM.
\newblock Low dimensional behavior of large systems of globally coupled
  oscillators.
\newblock Chaos. 2008;18(3):037113.
\newblock doi:{10.1063/1.2930766}.

\bibitem{OA09}
Ott E, Antonsen TM.
\newblock Long time evolution of phase oscillator systems.
\newblock Chaos. 2009;19(2):023117.
\newblock doi:{10.1063/1.3136851}.

\bibitem{OHA11}
Ott E, Hunt BR, Antonsen TM.
\newblock Comment on ``Long time evolution of phase oscillators systems''.
\newblock Chaos. 2011;21:025112.

\bibitem{Mat16}
Mattia M.
\newblock Low-dimensional firing rate dynamics of spiking neuron networks.
\newblock arXiv preprint arXiv:160908855. 2016;.

\bibitem{ALB+17}
Augustin M, Ladenbauer J, Baumann F, Obermayer K.
\newblock Low-dimensional spike rate models derived from networks of adaptive
  integrate-and-fire neurons: comparison and implementation.
\newblock PLOS Computational Biology. 2017;13(6).
\newblock doi:{10.1371/journal.pcbi.1005545}.

\bibitem{SDG17}
Schwalger T, Deger M, Gerstner W.
\newblock Towards a theory of cortical columns: From spiking neurons to
  interacting neural populations of finite size.
\newblock PLOS Computational Biology. 2017;13(4):1--63.
\newblock doi:{10.1371/journal.pcbi.1005507}.

\bibitem{PM16}
Paz\'o D, Montbri\'o E.
\newblock From Quasiperiodic Partial Synchronization to Collective Chaos in
  Populations of Inhibitory Neurons with Delay.
\newblock Phys Rev Lett. 2016;116:238101.
\newblock doi:{10.1103/PhysRevLett.116.238101}.

\bibitem{RP16}
Ratas I, Pyragas K.
\newblock Macroscopic self-oscillations and aging transition in a network of
  synaptically coupled quadratic integrate-and-fire neurons.
\newblock Phys Rev E. 2016;94:032215.
\newblock doi:{10.1103/PhysRevE.94.032215}.

\end{thebibliography}

\end{document}